\documentstyle[amsfonts,epsfig,portland]{aipproc}

\begin{document}
\rightline{\small Proceedings of the International Symposium}
\rightline{\small\it Capture Gamma-ray Spectroscopy and Related Topics}
\rightline{\small Santa Fe, New Mexico, August 29-September 3, 1999}
\vspace{-1.0pc}

\title{Electromagnetic response of light nuclei}

\author{Alberto Mengoni${}^{*} {}^{\dagger} {}^{\natural}$ 
and Takaharu Otsuka$^{\dagger} {}^{\natural}$}
\address{$^*$ENEA, Applied Physics Division,
Via Don Fiammelli 2 - 40129 Bologna, Italy\\
$^{\dagger}$The University of Tokyo, Department of Physics,
2-3-5 Hongo, Tokyo 113, Japan \\
$^{\natural}$RIKEN, 2-1 Hirosawa, Wako, Saitama 351-01, Japan}

\maketitle

\begin{abstract}
We show examples of neutron capture, photo dissociation
and Coulomb dissociation processes, relevant for studying
nuclear structure properties of some light nuclei.
In the case of the neutron capture, we show how 
interference effects among the direct and resonance
processes can be accounted for. The same interference effect
is shown to play an important role in the
photo dissociation of $^{9}{\rm Be}$, for
excitation energies just above
the neutron emission threshold. Finally,
the Coulomb dissociation of radioactive
projectiles is shown to provides basic
structure information on neutron rich
exotic systems. The example of the
$^{19}{\rm C}$ halo structure is shown.
\end{abstract}

\section*{Introduction}
Nuclear structure properties of light nuclei can be investigated by
studying their response to electromagnetic excitations.
One way to adopt this method is the traditional
slow and/or fast neutron capture $\gamma$-ray emission
spectroscopy. In addition, textbook examples of photo-dissociation
processes (e.g. $\gamma + d \rightarrow n + p$)
exist showing how the time-reversal 
reaction channel can also be a corresponding powerful 
technique.
Coulomb excitation processes have been traditionally
employed to study the structure properties of
nuclear bound states. More recently, a technique has
been developed in which the Coulomb excitation process
has been extended above the dissociation limit
(namely above some nuclear breakup channel). The
use of this technique became particularly attractive
with the availability of radioactive ion beams (RIBs).
In fact, using RIBs this method allows to do
spectroscopy of nuclei far from the stability line, 
up to the drip-lines.
We will present here examples of each one of these techniques
just mentioned, with particular emphasis on new developments
concerning the traditional approaches as well as 
with some very recent results obtained with the
up-to-date RIBs facilities.

\section*{Neutron capture}
For light nuclei, the neutron capture process
is often dominated by a direct radiative capture
process (DRC). This is a process in which the incident
neutron makes a direct transition into one
of the bound states of the residual nucleus
emitting one or more $\gamma$-rays, {\it without}
forming any intermediate compound state.
DRC models have been formulated in different
ways, but with, of course, the same physical
contents. To carry on the discussion on the
neutron capture as well as on the 
dissociation processes we need to give here 
one of these formulation.
We start from the wave functions $\Psi_{c}$ and $\Phi_{b}$ of the
continuum and of the bound-state, respectively given by
\begin{equation}
\Psi_{c}(r) = \displaystyle\sum_{l_c j_c} \\\
i^{l_c} \\\
\frac{\psi_{E_{c} l_c j_c}(r)}{r} \\\
[ [{\hat Y}_{l_c} \times {\hat \chi}_{s}]^{j_c} 
\times {\hat \chi}_{I}]^{J_c}.
\end{equation}
and
\begin{equation}
\Phi_{b}(r) = \displaystyle\sum_{l_b j_b} \\\
B_{j_b, J_b \alpha_b, I \alpha} \\\
\frac{\phi_{n_{b} l_b j_b}(r)}{r} \\\
[ [{\hat Y}_{l_b} \times {\hat \chi}_{s}]^{j_b} \times {\hat \chi}_{I}]^{J_b}
\end{equation}
Here, ${\hat Y}_{l}$ and ${\hat \chi}_{s}$ 
are angular and spin components of the wave functions,
the spin of the target nucleus is indicated by $I$ and all other
quantum numbers defining the continuum, bound and core state uniquely 
are indicated by $\alpha_c$, $\alpha_b$ and $\alpha$ respectively.
The $B$'s coefficients are the fractional parentage amplitudes
determined by the total number of particles occupying the
$n_b l_b j_b$ orbit, $N$. 
In shell-model standard notation\cite{BruGla77}
they are indicated by 
$< (j_b)^{N} J_{b} \alpha_{b} \mid \rbrace (j_b)^{N-1} I \alpha>$
and the spectroscopic factor for a given configuration
is simply given by $S = N \cdot B^{2}$.
The radial wave functions $\psi_{E_c l_c j_c}(r)$ and 
$\phi_{n_b l_b j_b}(r)$ are obtained by solving the
scattering and the bound state problem respectively,
in a given potential. 
The matrix elements for electric $L$-pole transitions are
$Q^{({\rm EL})}_{c \rightarrow b} = <\Psi_{c} 
\mid\mid \hat{T}^{\rm EL}\mid\mid
\Phi_{b}>$ and may be written as a product of three terms
\begin{equation}
Q^{({\rm EL})}_{c \rightarrow b} = {\cal I}_{c b} \cdot B_{b} \cdot A_{c b}
\end{equation}
with the electric transition operator 
$\hat{T}^{\rm EL}_{M} = r^{L} Y_{L M}$
and the radial overlap integral defined as
\begin{equation}
{\cal I}_{c b} = \displaystyle\int 
\psi_{E_c l_c j_c}^{*}(r) r^{L}  \phi_{n_b l_b j_b}(r) dr.
\end{equation}
The angular-spin coupling coefficient is indicated
by $A_{cb}$ and its definition can be found in the
reference\cite{Mex95}.
The capture cross section is given by
\begin{equation}
\sigma_{c \rightarrow b}^{({\rm EL})} =
\frac{8\pi(L+1)}{L[(2L+1)!!]^2}
\frac{k_{\gamma}^{2L+1}}{\hbar v} 
\frac{1}{2s+1} \frac{1}{2I+1}  \bar{e}^{2}_{\rm EL} 
\displaystyle\sum_{l_c j_c, l_b j_b} 
\mid Q^{({\rm EL})}_{c \rightarrow b} \mid^{2}
\end{equation}
where the sum is extended to all components of the
the continuum and bound state wave functions
compatible with the ${\rm EL}$ selection rules. 
Here, $k_{\gamma} = \epsilon_{\gamma}/\hbar c$
is the photon wave number corresponding to a 
transition energy $\epsilon_{\gamma}$ 
and $v$ the core-neutron 
relative velocity in the continuum.
$\bar{e}_{\rm EL}$ is the single-particle effective charge
of the neutron.

In the presence of a resonance state in the vicinity
of the continuum energy $E_c$, the capture cross section
contains a Breit-Wigner and an interference terms, 
in addition to the DRC term just shown
\begin{equation}
\sigma(E) = \sigma_{DRC}(E) + \sigma_{BW}(E) 
+ 2 [ \sigma_{DRC}(E) + \sigma_{BW}(E) ]^{1/2} \cos[\delta_{r}(E)]
\end{equation}
with
\begin{equation}
\sigma_{BW}(E) = \pi \lambda_{n}^{2} g_{J}
\frac{\Gamma_{n} \Gamma_{\gamma}}{(E_{r}-E)^{2} + 
\frac{1}{4} \Gamma^{2}}
\end{equation}
and
\begin{equation}
\delta_{r}(E) = \tan^{-1} [\frac{\Gamma(E)}{2(E-E_{r})}]. 
\end{equation}
Here, the resonance energy, the neutron, gamma and total widths
are indicated by $E_r, \Gamma_n$, $\Gamma_\gamma$ and $\Gamma$, 
respectively. The statistical weight factor is as usual
$g_J = (2J_b + 1)/(2I + 1)(2s + 1)$. The $\Gamma_\gamma$ is
intended to indicate the partial radiative width for a transition
leading to a specific bound state.

\begin{figure}[!t] 
\centerline{\epsfig{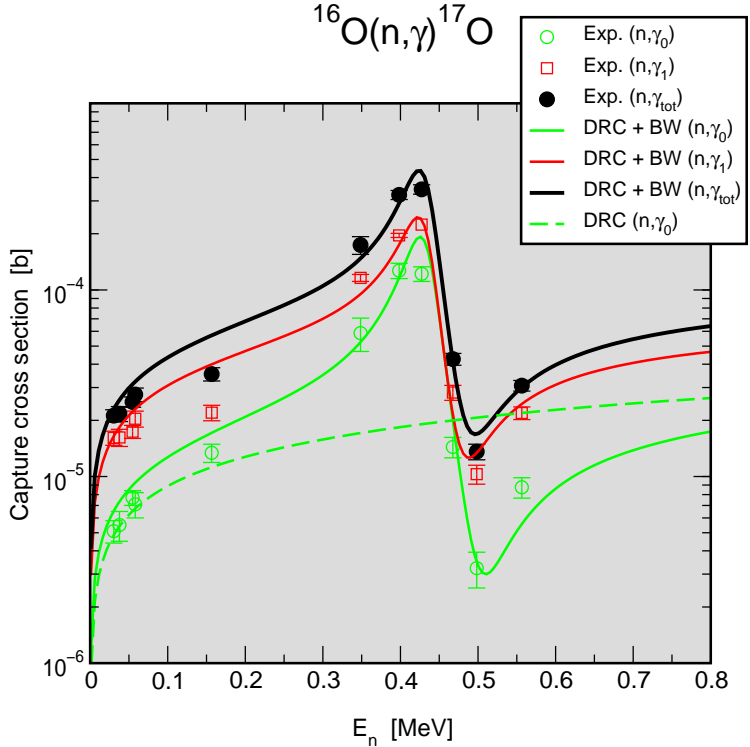}}
\caption{$^{16}{\rm O}(n,\gamma)^{17}{\rm O}$ cross section
for neutron energies around the 434 KeV neutron resonance. The
experimental values are from\protect\cite{Nagai99}. $(n,\gamma_{0})$ and
$(n,\gamma_{1})$ indicates the transition leading to the
ground and first excited state in $^{17}{\rm O}$, respectively.}
\label{fig1}
\end{figure}

A nice example of the onset of the resonance and interference
terms in the neutron capture cross section is provided
by the $^{16}{\rm O}(n,\gamma)^{17}{\rm O}$ reaction
for energies up to about 1 MeV. Here, the well known
$J^{\pi} = 3/2^{-}$ $p-$wave resonance at 434 KeV,
interferes with the smooth DRC $p-$wave component
as shown in Figure \ref{fig1}. 
Both the capture transitions leading 
to the $J^{\pi} = 5/2^{+}$ ground state
and to the $J^{\pi} = 1/2^{+}$ first excited state in $^{17}{\rm O}$ 
show typical interference patterns as a function of the
incident neutron energy. The calculation of the DRC component
was done using the $1d_{5/2}$ and $2s_{1/2}$ single-particle
wave functions calculated from a Woods-Saxon potential with
standard geometrical parameters $r_{0}=1.236$ fm, $d=0.62$ fm
and $V_{so}=7.5$ MeV. The well depth was adjusted to 
reproduce the experimental binding energies of the two states
($V_{0}=52.9$ and $V_{0}=54.5$ respectively for the
ground and first excited state). The spectroscopic
factors for both the bound states were taken to be unity.
The wave functions for the continuum were calculated using
the same potential parameters with a well-depth
of $V_{0}=54.5$ MeV. With a resonance energy $E_{r} = 434$ KeV
and neutron width $\Gamma_{n} = 45 $ KeV, the partial
gamma-ray widths necessary to reproduce the experimental
data were $\Gamma_{\gamma_0} = \Gamma_{\gamma_1} = 0.5 $ eV.
No other adjustable parameters were used in the calculation. 
The obtained total radiative width was therefore
$\Gamma_{\gamma} = 1.0 $ eV, 40\% smaller than the
BNL compiled value\cite{Mug81}.
\vspace{1.5pc}

\section*{Photo dissociation}
The time-reversal invariance of nuclear reactions provides
a relation between the capture and photo-neutron cross
section (detailed balance). 
In fact, once we define the reaction as
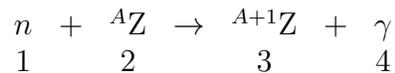
\begin{center}
$ \begin{array}[h]{c c c c c c c}
n & + &  {}^{A}{\rm Z} &  \rightarrow & {}^{A+1}{\rm Z} & + & \gamma \\
1 &   &    2           &              &        3        &   &   4    \\
\end{array} $
\end{center}
the relation connecting the direct and
inverse processes is simply given by
\begin{equation}
\sigma_{\gamma , n} \\\ = 
\frac{k_{n}^{2}}{k_{\gamma}^{2}} \\\
\frac{2J_{2}+1}{2J_{3}+1} \\\ 
\sigma_{n, \gamma}
\end{equation}
where $k_{n}$ is the wavenumber of the
$n + {}^{A}{\rm Z}$ relative motion.

\begin{figure}[!t] 
\centerline{\epsfig{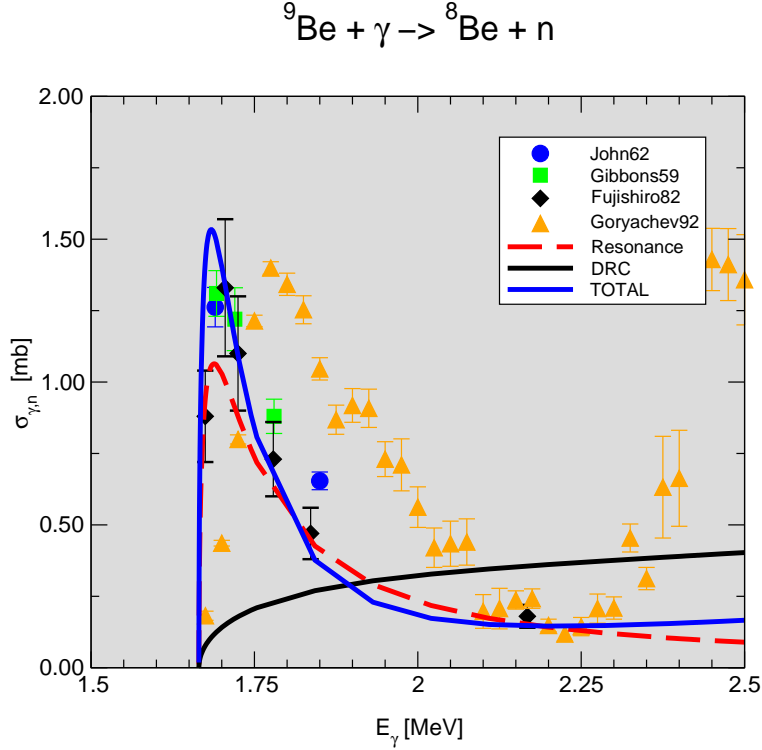}}
\caption{Photo dissociation of $^{9}{\rm Be}$ 
in the vicinity of the $^{8}{\rm Be} + n$ emission threshold. 
Experimental data are from the EXFOR compilation\protect\cite{EXFOR} 
(see there for references). The DRC, resonance and interference 
terms are calculated as described in the text.}
\label{fig2}
\end{figure}

An interesting example of photo-dissociation process
is given by the 
$\gamma + {}^{9}{\rm Be} \rightarrow {}^{8}{\rm Be} + n$
reaction. In fact, its time-reversal process, the neutron capture
by ${}^{8}{\rm Be}$, is supposed to play a crucial role
in the so-called $\alpha$-induced recombination process
during the post-collapse phase of a TypeII 
supernova\cite{Goerr95}.
The neutron capture reaction rate cannot be measured
directly because of the $\alpha$-decay of $^{8}{\rm Be}$.
However, the photo-dissociation experiment can be performed
and, indeed, some data are available.
From the nuclear structure point of view, this is an
interesting case as the first excited $J^{\pi}=1/2^{+}$ 
state of $^{9}{\rm Be}$, is located just above
the neutron separation threshold.
In view of the discussion we made above
concerning the interference between the DRC and
the resonance process, some interesting behavior 
of the photo-neutron cross section close to threshold 
can be expected. 
The structure properties of the $J^{\pi}=1/2^{+}$ 
state are also still quite uncertain.
Its most obvious property is that it is located
below the $J^{\pi}=5/2^{+}$ state. The lowering
of the $2s_{1/2}$ orbit in light systems (of
which the  $J^{\pi}=1/2^{+}$ state is supposed to
be mainly composed) is the subject or recent interesting 
investigations.

We have performed the calculation of the 
${}^{9}{\rm Be}(\gamma,n)^{8}{\rm Be}$ cross section
using a DRC model for the bound-to-continuum transition
plus resonance and interference contributions.
The results of this calculation is shown in
Figure \ref{fig2}.
The parameters and the technique used for the calculation
of the DRC component are similar to those employed
for the calculation of the neutron capture cross
section discussed above. Here, however, the 
${}^{9}{\rm Be}$ ground state has a
configuration of type 
\mbox{$\vert {}^{8}{\rm Be}(0^{+}) 
\times \nu(1p_{3/2});3/2^{-} \rangle$}. 
In addition, in this case we
made a shell model calculation using
the Cohen-Kurath residual interaction for
the p-shell model space and obtained
$S = 0.5$. The parameters of the
resonance state were taken from
the work of Barker\cite{Bark83}. 
These were
$E_{r}=73.4$ KeV, $\Gamma_{n}=255$ KeV and
$\Gamma_{\gamma}=0.53$ eV.
From the results shown in Figure \ref{fig2}
we can see that the peak observed just
above threshold is due to the presence
of the $J^{\pi}=1/2^{+}$ state. The
DRC component is rather smooth and an
interference effect seems to take place
reducing the cross section at 
$\epsilon_{\gamma} \approx 2$ MeV below the
DRC component.

\section*{Coulomb dissociation}
If a reaction is induced by real photons, the
above equation can be employed directly to relate
the direct and inverse channel. If, however, 
the photo-neutron process is to be induced by the
virtual photons provided by a heavy-$Z$ target,
like in for heavy-ion collisions, the process is
more complicated.

\begin{figure}[!t] 
\centerline{\epsfig{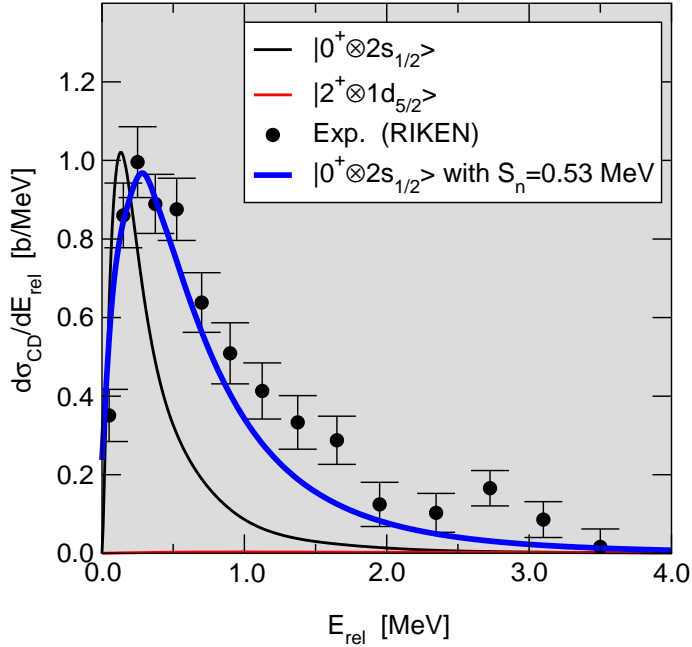}}
\caption{Coulomb dissociation cross section spectrum for 
${}^{19}{\rm C} \rightarrow {}^{18}{\rm C} + n$ at 67 A MeV 
incident energy as a function of the ${}^{18}{\rm C}-n$
relative energy.
The experimental data are from a recent RIKEN
experiment\protect\cite{Nakx99}. The solid thin line shows
the cross section as derived assuming a neutron separation
energy of 160 KeV. \mbox{$\vert {}^{18}{\rm C}(2^{+}) 
\times \nu(1d_{5/2});1/2^{+} \rangle$} configuration
contribution is negligible. The main configuration
is the \mbox{$\vert {}^{18}{\rm C}(0^{+}) 
\times \nu(2s_{1/2});1/2^{+} \rangle$} with a
spectroscopic factor of $S = 0.67$. See\protect\cite{Nakx99}
for details.}
\label{fig3}
\end{figure}

In the semi-classical limit (Coulomb trajectories) 
the cross section distribution for Coulomb dissociation 
is related to the photo-nuclear cross section by the relation
\begin{equation}
\frac{d\sigma_{cd}^{\rm EL}}{dE_{x}} = 
\displaystyle\int 2 \pi b db 
\frac{N_{\rm EL}(E_{i},E_{x},b)}{E_{x}}
\sigma_{\gamma,n}^{\rm EL} =
\frac{N_{\rm EL}(E_{i},E_{x})}{E_{x}}
\sigma_{\gamma,n}^{\rm EL} 
\end{equation}
where $E_{x} = S_{n} + E_{rel}$ is the excitation energy 
($E_{rel}$ is the ${}^{18}{\rm C} - n$ relative energy)
and $N_{\rm EL}$ the virtual-photon number 
of multipolarity ${\rm EL}$ at given
incident energy $E_{i}$ and impact parameter $b$.
Analytical expression for ${\rm EL}$ can be worked out
and can be found in the literature\cite{BerBau88}.

Many breakup experiments have been recently
performed using incident radioactive ion beams.
Examples includes 
$^{8}{\rm B} \rightarrow {}^{7}{\rm Be} + p$, 
$^{11}{\rm Li} \rightarrow {}^{9}{\rm Li} + 2n$,
$^{11}{\rm Be} \rightarrow {}^{10}{\rm Be} + n$, 
$^{14}{\rm Be} \rightarrow {}^{12}{\rm Be} + 2n$
and others. 
We report here on the results recently obtained in
RIKEN of the breakup of $^{19}{\rm C}$, the
heaviest of these loosely bound systems observed 
so far.
This case is interesting as the structure of
$^{19}{\rm C}$ was poorly known. The spin and
parity of the ground state was uncertain as
well as its neutron separation energy.
The details of the experiment can be found in
the reference\cite{Nakx99}. Here we show the
experimental results together with our
calculation in Figure \ref{fig3}. The
experiment was performed at $E_{i} = 67 $ A MeV.
From a comparison of the experimental data with
the theoretical calculation it has been concluded 
that: 1) the spin and parity of the $^{19}{\rm C}$ 
ground state is $1/2^{+}$, 2) the neutron
binding energy is $S_{n} = 0.53 \pm 0.13$ MeV.
Here we limit ourselves in observing that the
Coulomb dissociation spectrum is well reproduced
by the calculation based on a direct breakup
model in which the $^{19}{\rm C}$ 
ground state contains a large fraction 
($S = 0.67$) of the 
\mbox{$\vert {}^{18}{\rm C}(0^{+}) 
\times \nu(2s_{1/2});1/2^{+} \rangle$} configuration.

\acknowledgments{This work has been partially
supported by the European Union through its
{\it Science and Technology Programme in Japan}.
One of us (A.M.) would like to express his gratitude
to the Department for Physics of the University of
Tokyo and to RIKEN - Radiation Laboratory, 
for the kind hospitality offered during the period
in which the most part of this work has been made.
We have benefited from extensive and fruitful
discussion with M. Ishihara, T. Nakamura and Y. Nagai.}

\end{document}